# Characterizing Image Accessibility on Wikipedia across Languages

**Elisa Kreiss**
Stanford University

**Krishna Srinivasan**
Google Research

**Tiziano Piccardi**
Stanford University

**Jesus Adolfo Hermosillo**
Stanford University

**Cynthia Bennett**
Google Research

**Michael S. Bernstein**
Stanford University

**Meredith Ringel Morris**
Google Research

**Christopher Potts**
Stanford University

## Abstract

We make a first attempt to characterize image accessibility on Wikipedia across languages, present new experimental results that can inform efforts to assess description quality, and offer some strategies to improve Wikipedia's image accessibility.

## Introduction

Images are an essential component of the Wikipedia experience (Rama et al., 2022, Srinivasan et al., 2021). There are over 5 million distinct images on English-language Wikipedia alone, and images receive by far the highest engagement of all multimedia content. They especially enrich the user experience of shorter articles, arts, and biographies of less well-known people (Rama et al., 2022).

Despite their clear relevance to readers, Wikipedia's images are largely inaccessible to blind and low-vision (BLV) users. To experience these images online, BLV users have to rely on tools such as screen readers that read out descriptions provided in the image's HTML alt tag. Previous work suggests that only about 6% of images on English language Wikipedia contain associated alt descriptions, and many fewer are estimated to be useful (Kreiss et al., 2022a).

In this paper, we take stock of the state of Wikipedia's image accessibility across languages. Our investigation is based on the WIT dataset (Srinivasan et al., 2021), which contains 108 languages, including dominant and low-resource languages. Our analysis highlights that the lack of image accessibility on Wikipedia is a crosslingual problem. We additionally present the results of an experiment in which sighted participants rated descriptions for quality, building on results from Kreiss et al. 2022b for sighted and BLV users, and we suggest some strategies that could be used to improve image accessibility overall.

## Image-based Texts on Wikipedia

We analyze the WIT dataset, an August 2020 snapshot of Wikipedia with a focus on images and all associated texts (Srinivasan et al., 2021). WIT includes all articles from languages that have more than 12K images, and images that have Creative Commons licenses and at least one associated image-based text. As illustrated in Figure 1, the images are richly contextualized including the three main image-based text types present on Wikipedia: caption, alt description, and attribution description. Each of these fulfills a distinct purpose, which we summarize here:

An image's **caption** appears below the image in an article, visible to all readers. It provides information that supplements what is already apparent from the image. An image's **alt description** takes the image's place to make it nonvisually accessible. An **attribution description** is a generic summary text provided for any image on Wikimedia, which contains information that can be directly extracted from the image (commonly in alt descriptions), as well as more high level information, such as the date the image was taken (commonly in captions).

While alt descriptions are the texts used to provide image accessibility, previous work suggests that other image-based texts could help generate those descriptions at scale (Kreiss et al., 2022). We therefore assess the availability of all texts for Wikipedia, since they might be a valuable resource for potential mitigation strategies.

## Image Accessibility Across Languages

We start by providing a perspective on the overall coverage of the three image-based texts in each language. This is a measure of whether text is present and doesn't provide any insight into the proportion of useful texts. As presented in Figure 2, the coverage of images with different types of associated texts is fairly uniform across languages. Most images (about 90%) have attribution descriptions, which are only connected to the image and not to the article where the image appears. Captions cover the second largest number of images (about 50%), which is a conservative estimate since images from info boxes have text that fulfills the captioning purpose but are not captured in WIT. Alt descriptions are the least available, covering on average 10% of the images in a given language. Although these results are only based on the existence of the text and not yet quality, they already outline a universal image accessibility issue on Wikipedia.

One indication of alt description quality in a language is the number of unique descriptions. Informative descriptions can be expected to be quite specific to the image they describe, meaning that a high proportion of unique descriptions is an indication of more informative descriptions. In most languages, about 80% of all alt descriptions are unique. Noteworthy outliers among highly represented languages are German and French, with only about 65% unique descriptions, as well as Polish, Chechen, and Slovenian, with less than 10%. The high duplication rate in these cases could be an indication of a categorical issue in the description writing process in these languages. In fact, when following up on these results, it seems that the high duplication rate in German is due to extensive use of image galleries in articles, where the alt description is defined by the template engine to be the same as the caption.

While these numbers provide some cross-lingual upper-limit estimates of how many alt descriptions there are, estimating how many **useful** descriptions there are is much more challenging. We conducted an experiment on Prolific, where we showed participants sampled alt texts from Wikipedia, which they rated according to the criteria established by Kreiss et al. 2022b. Each of the 100 participants was introduced to the image accessibility problem and rated 4 out of the 30 descriptions of interest, which were presented within the available Wikipedia context (see Figure 1). The experiment, data and analysis are made available.[1]

Replicating previous work, we find that longer descriptions are more likely to be rated as high-quality. However, we don't find evidence for a generally applicable heuristic that would allow for a generalizable automatic estimate of how many descriptions are useful. There seems not to be a firm minimum length requirement, nor are descriptions that are repetitions from the main text uniformly considered bad. From the 30 descriptions tested, only two are consistently rated as good by annotators. About half of the descriptions are on average rated as rather not good (less than 3 out of 5 on the Likert scale). These results suggest that, even for English, the rate of overall alt text coverage might severely overestimate the rate of useful alt text, but automatically detecting and flagging low quality descriptions will require sophisticated methods guided by empirical research.

## Mitigation Strategies.

**Engaging Editors:** Wikipedia is written by a devoted community of volunteers who rely on norms and common practices to ensure growth and quality. Engaging them on issues of accessibility could have a powerful effect. For example, we estimate that more than 40% of the English Wikipedia articles marked by the community as Featured (best quality) have no alt texts for any of their images. Explicitly including accessibility as a quality dimension could help. There may be "accessibility by design" steps that would help as well. Accessibility could be enforced in the editing tool, by requiring contributors to write alt text for every new image before saving the content. Direct community engagement through cultural organizations[2] and dedicated events[3] is likely to provide opportunities for editors to focus on content that needs urgent attention and connect contributors with BLV people who can offer valuable feedback on their needs.

**Tools to support editing:** Some accessibility challenges clearly trace to avoidable limitations of current tools. First, browsers do not show alt texts by default, making missing alt text hard for sighted users to notice. Tools to surface articles and images without accessibility coverage would mitigate this, as would interfaces to help editors prioritize areas of need. Second, as our experiments show, writing good alt text is hard. Our tools must consider this aspect, providing clear instructions and, potentially, feedback on the quality of the text written. Third, although the general position of the Wikipedia community is to avoid adding any content written by generative models,[4] mixed human–AI systems may prove effective. These systems may include a human-supervised translation of the alt-text used in other languages for the same image or automatic captioning models that, despite shortcomings, may lower the participation barrier, since it is often easier to edit an existing text than it is to write one from scratch.

**Successful communities**: We may be able to find areas of Wikipedia that are unusually successful in achieving good coverage for accessibility descriptions, and such communities could teach us important lessons.

---

[1] https://github.com/elisakreiss/wikiworkshop2023_imgaccessibility

[2] https://en.wikipedia.org/wiki/Wikipedia:GLAM/Smithsonian_Institution/Accessibility

[3] https://en.wikipedia.org/wiki/Edit-a-thon

[4] https://en.wikipedia.org/wiki/Wikipedia:Using_neural_network_language_models_on_Wikipedia

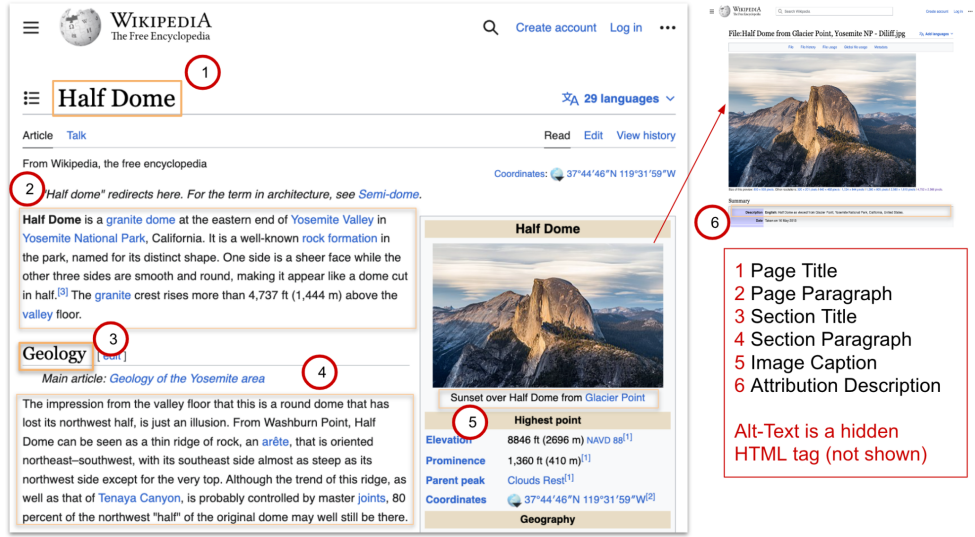

Figure 1: Overview of the variety of texts associated with an image on Wikipedia.
Wikipedia page for Half Dome, Yosemite National Park, CA: https://en.wikipedia.org/wiki/Half_Dome

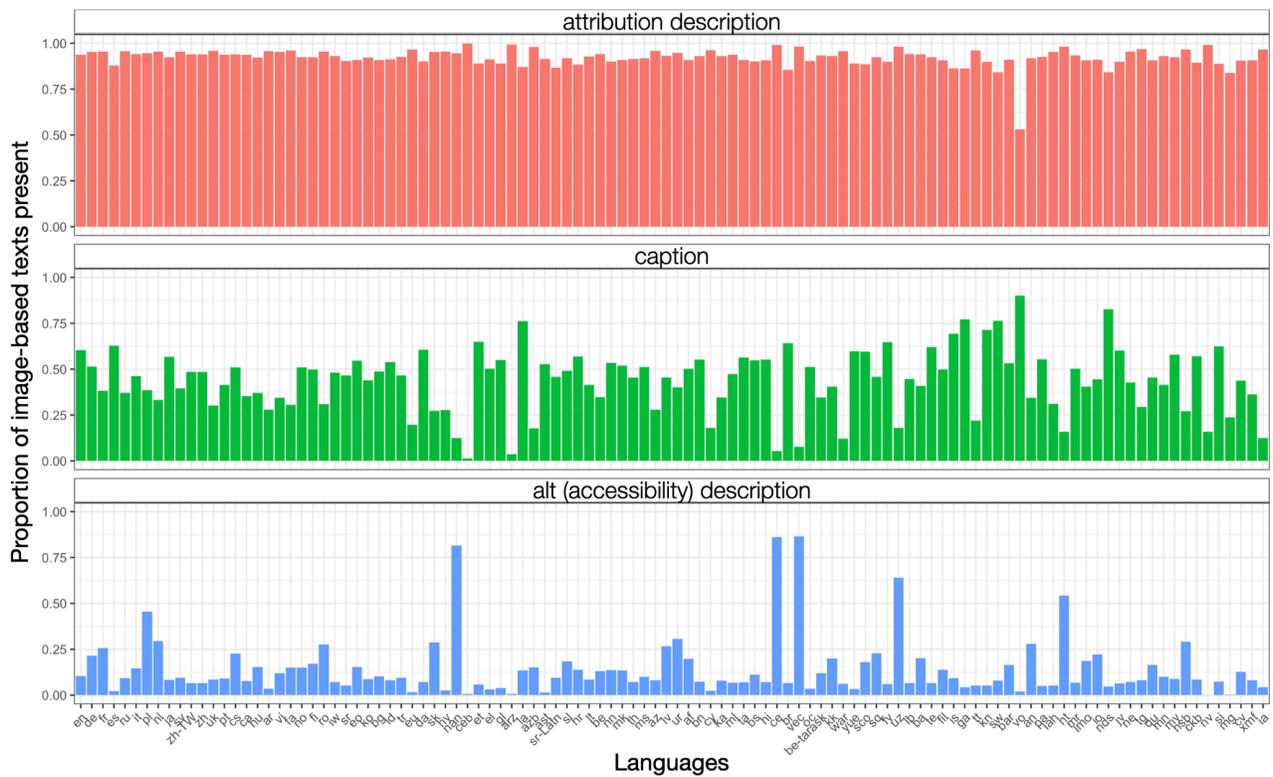

Figure 2: Proportion of the three distinct image-based texts for each image on Wikipedia across 108 languages, sorted by the number of images present for each language. While the coverage of alt descriptions is uniformly low, there is still remarkable variance between languages that is informative for understanding editing practices in different communities. English, the language with the most articles and images, only achieves average coverage (11%) whereas highly represented languages German (de, 20%) and French (fr, 25%) show significantly higher coverage. Spanish (es) is an outlier with many images but a remarkably small number of alt descriptions, suggesting a categorical lack of norms around alt description writing. However, among the whole set of languages, Venetian (vec), Chechen (ce), Southern Min (nan), Uzbek (uz) and Haitian (ht) stand out, since more images in these languages have alt descriptions than not. It's worth exploring which editing strategies might promote this and whether those methods should and can be adopted in the larger Wikipedia community.